# Polylithiated (OLi$_2$) functionalized graphane as a potential hydrogen storage material.

Tanveer. Hussain,[1] Tuhina. Adit. Maark,[1] Abir. De Sarkar,[1,2] and Rajeev. Ahuja [1,2]

[1]Condensed Matter Theory Group, Department of Physics and Astronomy, Box 516, Uppsala University S-75120 Uppsala, Sweden, EU

[2]Applied Materials Physics, Department of Materials and Engineering, Royal Institute of Technology (KTH) S-100 44 Stockholm, Sweden, EU

Hydrogen storage capacity, stability, bonding mechanism and the electronic structure of polylithiated molecules (OLi$_2$) functionalized graphane (CH) has been studied by means of first principle density functional theory (DFT). Molecular dynamics (MD) have confirmed the stability, while Bader charge analysis describe the bonding mechanism of OLi$_2$ with CH. The binding energy of OLi$_2$ on CH sheet has been found to be large enough to ensure its uniform distribution without any clustering. It has been found that each OLi$_2$ unit can adsorb up to six H$_2$ molecules resulting into a storage capacity of 12.90 wt% with adsorption energies within the range of practical H$_2$ storage application.

## 1. Introduction

The consumption of energy is increasing at a rapid pace and is predicted to be almost doubled over the next few decades. The current resources of energy are also decreasing with each passing day. The CO$_2$ emission caused by the extensive use of fossil fuels results in global warming and leaves devastating effects on atmosphere. So, there is a strong need for alternate sources of energy, which are safe, efficient, abundantly available, and environment friendly.[1] Hydrogen could be one of the best available choices as promising energy carrier due its abundant availability, highest energy density, environment friendliness and low cost.[2-5] But the gaseous nature of hydrogen and the unavailability of the storage media for practical applications, restricts its use as a energy carrier in fuel cell. Different storage media were considered for efficient $H_2$ storage in recent past and carbon based nanostructures are considered to be the most promising materials.[6-9]

Along with the other countless applications, carbonaceous nanomaterials including fullerenes, carbon nanotubes (CNTs), graphene etc. have extensively been used for energy applications, especially hydrogen storage purposes.[10-12] The greatest advantages of using carbon based materials are their light weight and low cost.

Carbon based nanostructures in pure form are too inert to be used in most of the technological applications unless they are functionalized with foreign elements. There exist a variety of elements for functionalization.

Transition metals have been the subject of many studies as dopant element of carbon nanostructures for the potential $H_2$ storage materials because of their strong binding to the substrates.[13-15] But the large values of their cohesive energies and higher atomic weight results into the cluster formation and low weight percentage of $H_2$ storage respectively. So, light element having low cohesive energies could be useful to avoid clustering and to have uniform coverage on carbon surfaces.

Alkali metals can be a good dopant on carbon nanostructures owing to its low cohesive energy ($E_{coh}$ ~1eV)[8] and uniform distribution on surface.

Graphene, having sp$^2$ bonding and possessing unusual electrical, unique mechanical, and extra ordinary optical properties is the most important member of carbon nanostructure family. It has opened up many windows right after its experimental isolation.[16]

In recent past graphene has been the subject of many studies regarding potential medium of H$_2$ storage. Ataca et al.[17] have found the calcium doped graphene as promising H$_2$ storage material reaching a storage capacity of 8.4 wt%. They also report the uniform distribution of Ca atoms on graphene sheet without cluster formation. They found Mg and Be unsuitable because unstable structures formed by these elements on graphene. Reunchan and Jhi [18] studied the interaction of different metals on porous graphene for H$_2$ storage. By means of first principle calculations they calculated the binding energies of different metal atoms and interaction of H$_2$ with different metal-doped porous graphene has been reported. Zhou et al. [19] investigated the adsorption of metal atoms on graphene under the application of strain. They also reported the improvement of metal-graphene binding and increased H$_2$ storage capacity under applied strain. Though there are different ways to enhance the binding of metal adatoms on graphene sheet, but still there has been issues in prospective to the strong metal-graphene binding strength and the possibilities for adatom-adatom clustering.

Beside graphene there is another interesting member of carbon nanostructure family called *graphane* or hydrogenated graphene. The exposure of graphene sheet to H$_2$ plasma results in the attachment of hydrogen atoms on each carbon atom on both sides of the sheet alternately giving rise to a crumbled structure in contrast to a planar graphene.

It has been verified both theoretically and experimentally. [20,21]

The advantage of using *graphane* as substrate to bind metal adatoms for storing hydrogen is the strong metal-*graphane* bonding. There are few studies describing the strong *graphane*- metal interaction and its usefulness in $H_2$ storage applications.[22,23] Along with the importance of adatom-substrate binding, the interaction of $H_2$ with the adatom (metal) is also very important for a good storage material. Lighter elements (Li, Na, Be, Mg etc.) prefer to bind the $H_2$ through electrostatic and polarization interactions.[24] For any storage media the optimum value of $H_2$ binding energy with the adsorbent should be around 0.15 eV.[25] This value ensures $H_2$ to be adsorbed and desorbed at ambient conditions.

The molecules having high density of lithium atoms (Li) are termed as polylithiated molecules. This class of species includes $CLi_n$ and $OLi_m$ (n=3-5 and m=1-4). There has been many theoretical as well as experimental studies predicting and verifying the existence of such Li rich species.[26-29] The polar nature of bonds in O-Li leave a significant amount of charge on Li atoms which in turn can polarize and adsorb $H_2$ molecules, thus resulting in a good storage of $H_2$ molecules.

In this work we have investigated the $H_2$ storage properties of $OLi_2$. Clustering between $OLi_2$ species on the surface may affect the reversibility of hydrogen storage and release. Consequently, to avoid such clustering we have studied the feasibility of attaching these species on both sides of the graphane sheet by removing two hydrogen atoms. The binding energy of $OLi_2$ unit on graphane has been calculated. We have found that at the most three $H_2$ molecules can be adsorbed on each Li atom resulting in a storage capacity of 12.90 wt% with binding (0.15-0.25 eV) feasible for practical use which is well above the target of 5.5 wt% set by Department of Energy (DOE) for on-board $H_2$ storage to be achieved by 2017.[30]

## 2. Computational method

The total energy calculations and structural relaxations in this study have been done by using VASP code based on density functional theory.[31-33] By using local density approximation (LDA) one can describe the covalent type of interaction in a better way but LDA is known to overestimation in energies. So, for reliable results, we have employed both generalized gradient approximation (GGA) and LDA.[34,35] For a better treatment of our weakly interacting systems ($H_2$ molecules adsorbed on Li atoms), we have calculated the van der Waals corrected interaction energies using the semi empirical correction of Grimme[36] as available with VASP .The Molecular dynamics calculations has also been carried out to confirm the binding stability of $OLi_2$ on *graphane* sheet up to 400 K by using projector-augmented wave approach as described in VASP.[37]

The unit cell of *graphane* have four atoms (2C, 2H), we have taken a 2x2x1 super cell containing 8 carbon and 8 hydrogen atoms for this study. A vacuum space of 15Å has been inserted in (001) direction perpendicular to the sheet to avoid the interaction among graphane sheets due to the periodicity. The Brillouin zone is sampled by 10x10x1 mesh points in k-Space based on Monkhorst-Pack scheme. For obtaining the density of states (DOS) we have engaged the tetrahedron method along with a 17x17x1 k-points mesh.[37] All the structures were fully relaxed until the force acting on each ion is less than 0.005 eV/Å.

## 3. Results and discussion

First of all, the geometry of pure graphane is discussed briefly. Fig.1 (a, b) shows the side and top view of the optimized structure of graphane. When the structure is fully optimized the C-C and C-H bonds are 1.53Å and 1.12Å respectively which is in good agreement with the previous studies.[20] Now two out of the eight hydrogen atoms on a graphane sheet are substituted with $OLi_2$ molecules, one each from (+Z) and (−Z) directions. This will result into a 25% doping concentration of $OLi_2$. Fig. 2a shows the optimized geometry of $C_8H_6O_2Li_4$ ($CHOLi_2$) respectively. The C-O and O-Li distances after full optimizations are 1.42Å and 1.78Å respectively. The calculation of binding energy ($\Delta E_b$) of $OLi_2$ molecule to graphane sheet is important to ensure its uniform distribution on the sheet. The following relation can be used to calculate $\Delta E_b$

$$\Delta E_b = E\,(CHOLi_2) - E\,(CH) - E\,(OLi_2) \quad (1)$$

Where E ($CHOLi_2$), E (CH) and E ($OLi_2$) are the total energies of $CHOLi_2$ ($C_8H_6O_2Li_4$), CH ($C_8H_6$) and $OLi_2$ ($O_2Li_4$) molecules respectively. The calculated $\Delta E_b$ of $OLi_2$ to CH sheet in this case is 2.60 eV, which is high enough to ensure the stability of the structure. It is important to mention here that the $OLi_2$ unit binds to the CH sheet through O-C bond. The stability of $OLi_2$ on CH sheet has been further confirmed by the application of molecular dynamics simulations. For this purpose the Nose-thermostat algorithm at 400 K with 1 fs time has been used. The structure of $CHOLi_2$ remains stable even after 4 ps, which confirms the stability of the structure. The MD results are also shown in Fig. 3.

In order to investigate the type of bonding between C-O and Li-O in in our designed structure $CHOLi_2$, we have performed the Bader charge analysis[38]. The electronegativity of O is much higher than that of C and Li. So, the electric charge is transferred to O from both C and Li atoms. After a careful analysis, it has been found that the each of C atom which is directly bonded with the O atom transfers ~ 0.90 e charge to the O atoms. At the same time each Li which is directly bonded with the O atoms in $OLi_2$ species donate ~ 0.97 electronic charge to the O atoms on either side of the CH sheet. In this way, O atoms acquire negative charge, while Li atoms gather positive charge. This high amount of charge on each Li atom facilitates the electrostatic attraction of $H_2$ molecules to be adsorbed on Li atoms. Some of the H atoms, which are in the close vicinity of O atoms, also attain a small amount of charge. Due to this charge transfer from C to O an ionic bond appear between them. Furthermore, the total and partial density of states of $OLi_2$ functionalized graphane has been plotted and analyzed. In case of pure graphane, an insulating behavior with a wide band gap of 3.55 eV has been reported.[20] In case of $CHOLi_2$, the total and partial density of states in shown in Fig. 4. The broadened resonances in the density of states (DOS) plotted in Fig. 4 signify a strong mixing between the states of the constituent atoms. The DOS at -3 eV is mainly contributed by the $C_O$ (2p) and O (2p) and their hybridization. The contribution of Li-2s and $C_H$ (2p) at this energy appears to be much smaller. The adjacent peak at -3.5 arises chiefly due to the mixing between the Li-2s and O-2p states. The DOS at -5.5 eV shows a mixture of all the decomposed states shown in Fig. 3; however, it is contributed comparatively largely by $C_H$ (2p). Similarly, the DOS at -6.5 eV is a mixture of all the states shown in pDOS; yet, it shows a relatively greater contribution from $C_O$ (2p) and O(2p) states.

In the end, the $H_2$ storage capacity of our designed material $CHOLi_2$ is discussed. As the Li atoms bonded with the O atoms on the both side of CH sheet acquired fractional positive charge, so the $H_2$ molecules can be easily physisorbed around them. $H_2$ molecules are introduced around each $Li^+$ ion in stepwise manner. The mechanism of the $H_2$ adsorption on $Li^+$ ions can be explained by considering the fact that $Li^+$ ions polarizes the $H_2$ molecules, which results into the binding of $H_2$ with $Li^+$. Van der Waal's forces are considered to be responsible for the attraction between the $Li^+$ ions and $H_2$ molecules. In order to have maximum storage capacity, the $H_2$ molecules should bind to Li$^+$ ions at physisorption distance and maintain a reasonable distance within them so that the repulsion among them can be avoided. We have found that at the most 3H$_2$ molecules can be adsorbed on each Li$^+$ ion in CHOLi$_2$ system. This will results in a very high storage capacity of 12.90 wt%, which is well beyond the DOE target to be attained till 2017. Fig. 2 (b, c, d) shows the optimized geometry of CHOLi$_2$ with H$_2$ molecules physisorbed on it. The adsorption energy ΔE$_{ads}$ of H$_2$ molecules can be calculated by

**ΔE$_{ads}$ = E {(CHOLi$_2$+nH$_2$) – E (CHOLi$_2$) – E (H$_2$)}/n     (2)**

Where **ΔE** (n) is the adsorption energy of the nth H$_2$ molecule adsorbed on the CHOLi$_2$ sheet, E (CHOLi$_2$) is the energy of CHOLi$_2$ sheet without H$_2$ molecule and E (H$_2$) is the energy of a single H$_2$ molecule. Table.1 shows the complete results describing the adsorption energies of ΔE$_{ads}$ (eV) of H$_2$ molecules adsorbed on CHOLi$_2$, and the average H-H bond length Δd (Å). For reliable results, and to avoid the overestimation of LDA and underestimation of GGA, we have also employed the van der Waal's corrected dispersion term in our calculations. The consistency in the values of ΔE$_{ads}$, and Δd is clear from the Table, regardless of the XC functional.

## 4. Conclusions

By using first-principle calculations, we have predicted that the polylithiated (OLi$_2$) functionalized graphane can serve as a fascinating material for high capacity H$_2$ storage. The structure of CHOLi$_2$ is stable and the large value of binding energy of OLi$_2$ on CH sheet will induce its uniform distribution on the sheet. Bader charge analysis indicates the polar nature of C-O and Li-O bonds. It has been found that a partially charged Li$^+$ ion on each OLi$_2$ molecule can adsorb up to six H$_2$ molecules resulting in a very high storage capacity of 12.90 wt%. The average adsorption energies of $H_2$ molecules have been found to be within the range of practical applications.

**Acknowledgements**

TH is thankful to higher education commission of Pakistan for doctoral fellowship. ADS and TAM are grateful to the Wenner-Gren Foundation and FORMAS for postdoctoral Fellowship. RA acknowledges FORMAS, SWECO and Wenner-Gren Foundation for financial support. SNIC and UPPMAX are acknowledged for computing time

Table 1. Calculated adsorption energies $\Delta E_{ads}$ (eV) per $H_2$, of the nth (n=4,8,12) adsorbed on $CHOLi_2$, and the average H-H bond length $\Delta d$ (Å) by using LDA, GGA and van der Waal's interaction included calculations.

| No. of $H_2$ Molecules | LDA | | GGA | | vDWaal's | |
|---|---|---|---|---|---|---|
| | $\Delta E_{ads}$ (eV) | $\Delta d$ (Å) | $\Delta E_{ads}$ (eV) | $\Delta d$ (Å) | $\Delta E_{ads}$ (eV) | $\Delta d$ (Å) |
| 4. | 0.450 | 0.830 | 0.155 | 0.765 | 0.200 | 0.770 |
| 8. | 0.272 | 0.820 | 0.110 | 0.790 | 0.150 | 0.792 |
| 12. | 0.181 | 0.793 | 0.108 | 0.773 | 0.175 | 0.772 |

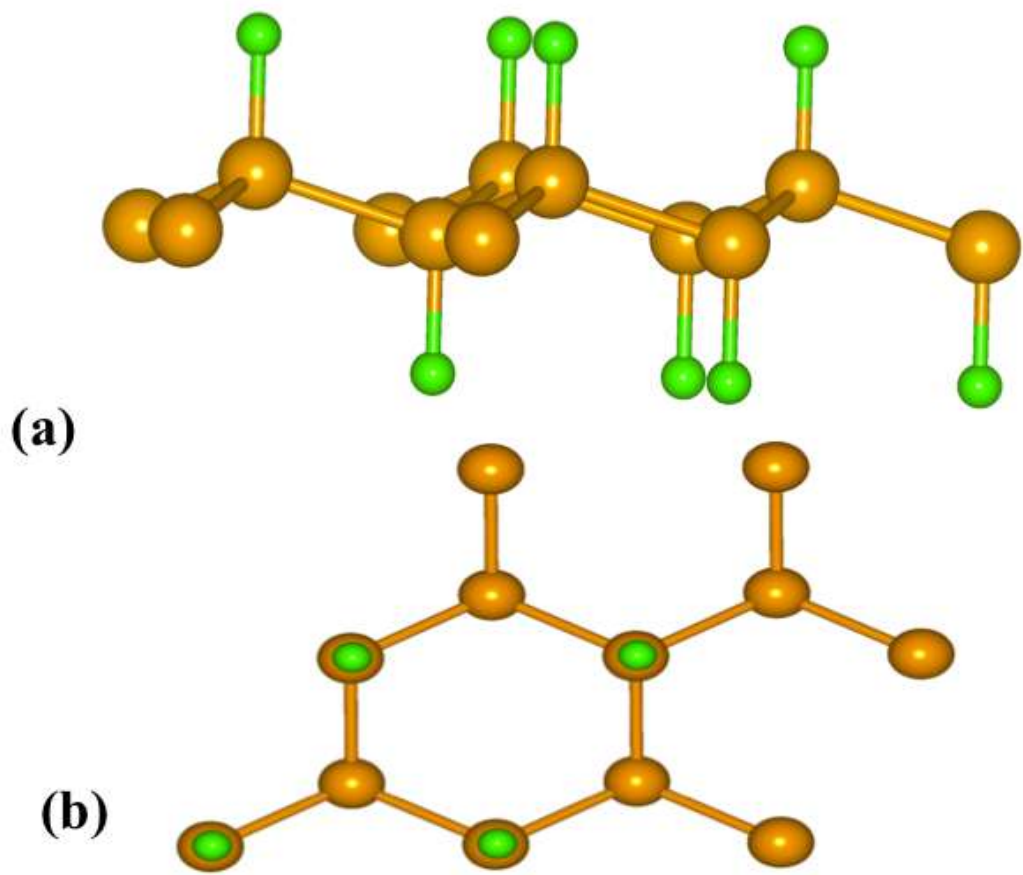

Fig. 1. Optimized structures of (a) side and (b) top view of pure CH. Carbon and hydrogen atoms are shown in yellow and green spheres respectively.

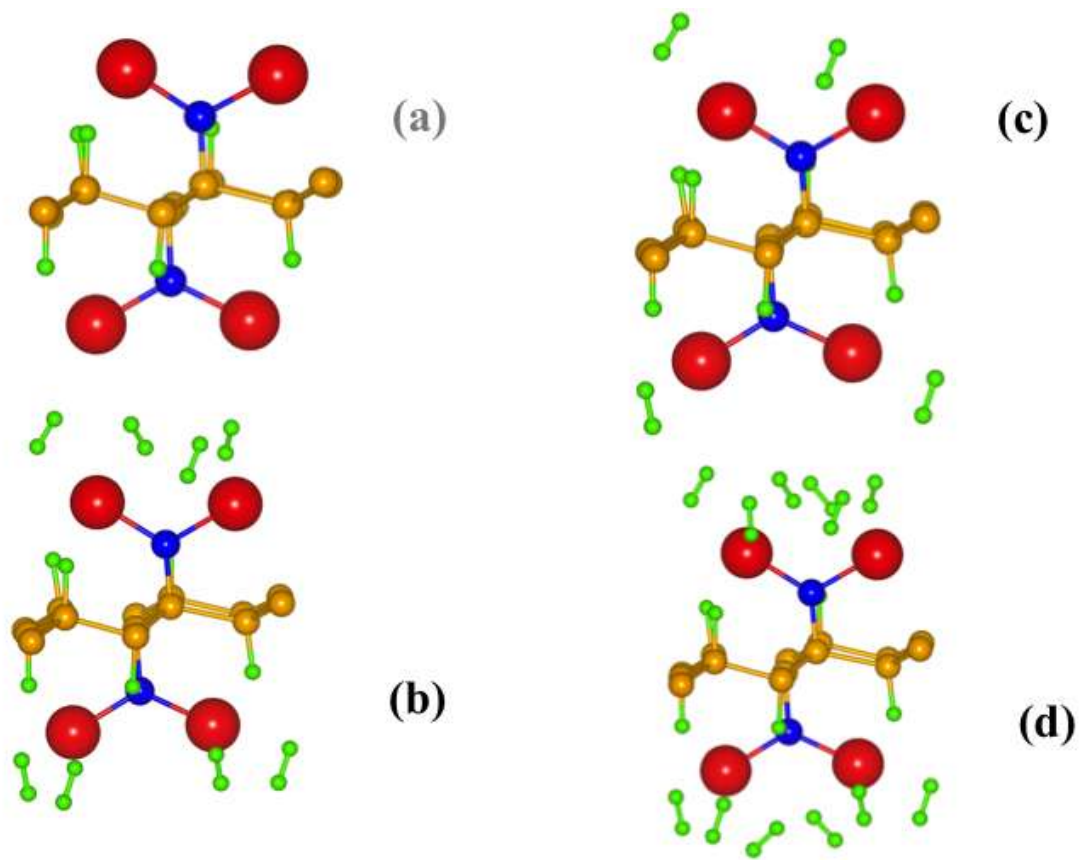

Fig.2. Optimized structures of (a) $CHOLi_2$, (b) $CHOLi_2 + 4H_2$ (c) $CHOLi_2 + 8H_2$ and (d) $CHOLi_2+12H_2$. Carbon, hydrogen, oxygen and lithium atoms are shown in yellow, green, blue and red spheres respectively.

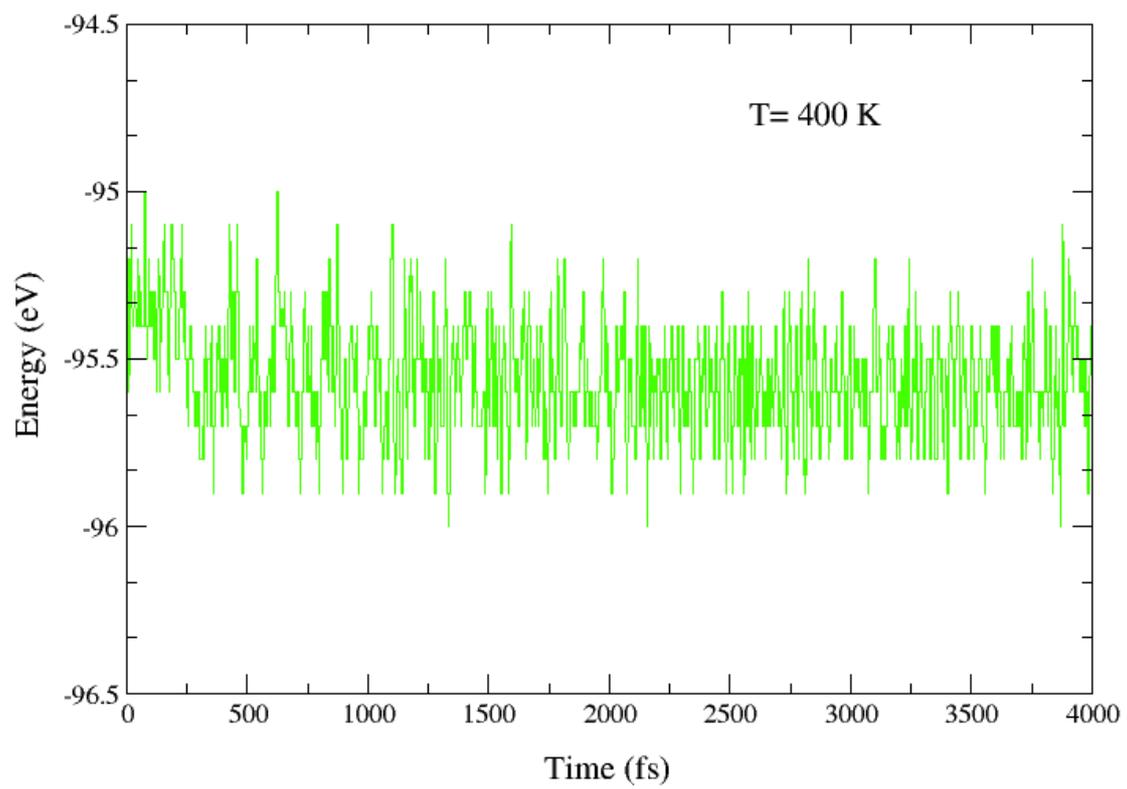

Fig. 3. Change of energy with time obtained from molecular dynamics simulation of $CHOLi_2$ sheet.

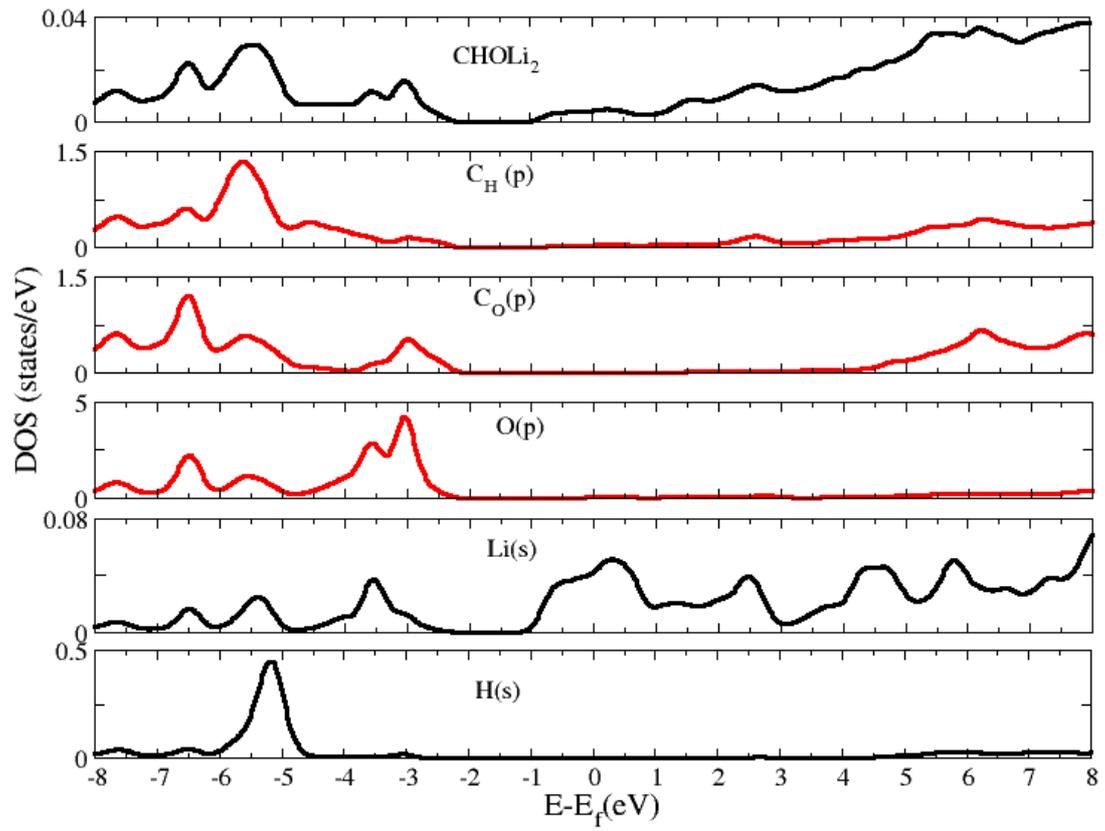

Fig. 4. Shows the total and partial density of states (DOS) of $CHOLi_2$ system. Fermi level sets to 0.